\documentclass{aa}  

\usepackage{graphicx}
\usepackage{txfonts}
\usepackage[pdfpagelabels=false]{hyperref}
\hypersetup{colorlinks=true,linkcolor=blue,citecolor=blue,filecolor=blue,urlcolor=blue,}
\makeatletter
\renewcommand*\aa@pageof{, page \thepage{} of \pageref*{LastPage}}
\makeatother

\usepackage{graphicx}	
\usepackage{amsmath}	
\usepackage{amssymb}	
\usepackage{multirow}   
\usepackage{array}      %
\usepackage{mathtools}
\usepackage[justification = centering]{subcaption}

\graphicspath{{./Figures/}}

\newcolumntype{P}[1]{>{\centering\arraybackslash}p{#1}}

\usepackage{xcolor}
\definecolor{green_comm}{RGB}{0,160,0}

\newcommand{\abs}{\vert}

\newcommand{\Hy}{\ion{H}{}}
\newcommand{\HI}{\ion{H}{i}}
\newcommand{\HII}{\ion{H}{ii}}
\newcommand{\He}{\ion{He}{}}
\newcommand{\HeI}{\ion{He}{i}}
\newcommand{\HeII}{\ion{He}{ii}}
\newcommand{\HeIII}{\ion{He}{iii}}

\newcommand{\fthr}{F_{\rm XUV}^{\rm thr}}

\newcommand{\gtsima}{$\; \buildrel > \over \sim \;$}
\newcommand{\ltsima}{$\; \buildrel < \over \sim \;$}
\newcommand{\prosima}{$\; \buildrel \propto \over \sim \;$}
\newcommand{\gsim}{\lower.5ex\hbox{\consistegtsima}}
\newcommand{\lsim}{\lower.5ex\hbox{\ltsima}}
\newcommand{\simgt}{\lower.5ex\hbox{\gtsima}}
\newcommand{\simlt}{\lower.5ex\hbox{\ltsima}}
\newcommand{\simpr}{\lower.5ex\hbox{\prosima}}

\newcommand{\Kphi}{K\phi_{\rm p}}
\newcommand{\flux}{F_{\rm XUV}}
\newcommand{\phired}{\phi_{\rm red}}
\newcommand{\frho}{\flux/\rho_p}
\newcommand{\etaeff}{\eta_{\rm eff}}

\begin{document}

\title{Irradiation-driven escape of primordial planetary atmospheres II. Evaporation efficiency of sub-Neptunes through hot Jupiters}

\author{Andrea Caldiroli\inst{1}
        \and Francesco Haardt\inst{2,3,4}
        \and Elena Gallo\inst{5} 
        \and Riccardo Spinelli\inst{2,4}
        \and Isaac Malsky\inst{5}
        \and Emily Rauscher\inst{5}}

\institute{
        Fakult\"at f\"ur Mathematik, Universit\"at Wien, Oskar-Morgenstern-Platz 1, A-1090 Wien, Austria
    \and 
        Dipartimento di Scienza e Alta Tecnologia, Universit\`a degli Studi dell'Insubria, via Valleggio 11, I-22100 Como, Italy
    \and
        INFN, Sezione Milano-Bicocca,P.za della Scienza 3, I-20126 Milano, Italy
    \and
        INAF, Osservatorio Astronomico di Brera, Via E. Bianchi 46, I-23807 Merate, Italy
    \and
        Department of Astronomy, University of Michigan, 1085 S University, Ann Arbor, Michigan 48109, USA
}


\abstract{
Making use of the publicly available 1D photoionization hydrodynamics code ATES we set out to investigate the combined effects of specific planetary gravitational potential energy ($\phi_p\equiv GM_p/R_p$) and stellar X-ray and Extreme Ultraviolet (XUV) irradiation ($F_{\rm XUV}$) on the evaporation efficiency ($\eta$) of moderately-to-highly irradiated gaseous planets, from sub-Neptunes through hot Jupiters. We show that the (known) existence of a threshold potential above which energy-limited thermal escape (i.e., $\eta\simeq 1$) is unattainable can be inferred analytically, by means of a balance between the ion binding energy and the volume-averaged mean excess energy. For $\log\phi_p\simgt \log\phi_p^{\rm thr}\approx [12.9-13.2]$ (in cgs units), most of the energy absorption occurs within a region where the average kinetic energy acquired by the ions through photo-electron collisions is insufficient for escape. This causes the evaporation efficiency to plummet with increasing $\phi_p$, by up to 4 orders of magnitude below the energy-limited value.
Whether or not planets with $\phi_p\simlt \phi_p^{\rm thr}$ exhibit energy-limited outflows is primarily regulated by the stellar irradiation level. Specifically, for low-gravity planets, above $\fthr \simeq 10^{4-5}$~erg~cm$^{-2}$~s$^{-1}$, Ly$\alpha$ losses overtake adiabatic and advective cooling and the evaporation efficiency of low-gravity planets drops below the energy-limited approximation, albeit remaining largely independent of $\phi_p$.
Further, we show that whereas $\eta$ increases as $F_{\rm XUV}$ increases for planets above $\phi^{\rm thr}_p$, the opposite is true for low-gravity planets (i.e., for sub-Neptunes). This behavior can be understood by examining the relative fractional contributions of advective and radiative losses as a function of atmospheric temperature. This novel framework enables a reliable, physically motivated prediction of the expected evaporation efficiency for a given planetary system; an analytical approximation of the best-fitting $\eta$ is given in the appendix. 
}


\titlerunning{Evaporation efficiency}
\maketitle
%

\section{Introduction}
\label{sec:intro} 

Several astrophysical phenomena are sustained by the conversion of gravitational potential energy of inflowing or outflowing matter into electromagnetic radiation, or vice versa. 
In this paper, we are concerned with the specific case where the primary energy source is in the form of stellar photo-ionizing radiation. The energy conversion--into heat--occurs within the atmosphere of a planet, where the ensuing thermal pressure gradient lowers the gas gravitational potential energy enough to drive an outflow \citep[e.g.,][amongst the seminal works]{Lammer2003,Lecavalier2004,Yelle2004,Tian2005,Erkaev2007,Koskinen2007,MC}. In this case, simple energy conservation balance yields ${GM_p\dot M}/{R_p}\simeq \pi R_{\rm XUV}^2 F_{\rm XUV}$, where $M_p$ and $R_p$ are the planet mass and radius, $F_{\rm XUV}$ is the photo-ionizing\footnote{For the purpose of this work, XUV denotes the energy range between 13.6 eV and 12.4 keV; the assumed spectral shape is summarized in \S\ref{sec:fluxappendix} and discussed in more detail in Paper I of this series \citep{ates1}.} stellar flux at the planet orbital distance, $R_{\rm XUV}$ denotes the distance (from the planet) at which most of the stellar flux is absorbed, and $\dot M$ is the resulting mass outflow rate. 
It is interesting to note that the same equation holds in classic accretion theory, where the dissipation of gravitational potential energy of the accreting gas, which sinks in at a rate $\dot{M}$, is partly converted into heat and generates an outward flux of radiation. Similar to the case of accretion flows, where much theoretical effort has gone into the determination of the radiative efficiency, bracketing the range of atmospheric outflow efficiencies remains a topic of active investigation \citep{lecavalier07,lammer09,Sanz-Forcada2011,Owen10,Owen2012,Erkaev2013,Shematovich2014,Chadney2015,khodachenko15,tripathi15,Salz2016a,Salz2016b,owenalvarez,Erkaev2016,kuby18a,Debrecht2019,odert20,kuby21,AllanVidotto2019,Vidotto2020}.\\
\indent The above energy conversion equation closely resembles the so-called energy-limited flux of escaping particles that was first derived by \citet{Watson1981} and applied to the specific case of Earth and Venus. 
The foundation of an energy-limited mass loss hinges on the role played by thermal conduction in limiting the radial extent of the thermosphere and thus the amount of stellar energy that is absorbed. Nevertheless, as noted by \cite{Owen2019}, the original connotation has been all but lost in the more recent literature. Rather, the energy-limited approximation is typically invoked to signify a theoretical maximum outflow rate which can be expected when a {dominant} fraction of the absorbed stellar radiation is converted into adiabatic expansion.

In practice, several unknowns hinder a straightforward estimate of this maximum $\dot{M}$. 
Firstly, a non negligible fraction of the stellar radiation that is absorbed by the atmosphere may go into exciting and ionizing its constituents. More broadly, radiative cooling competes with and may even offset adiabatic cooling. Second, the value of $R_{\rm XUV}$ cannot be readily estimated from first principles. Our ignorance of either parameter can be conveniently encapsulated in an efficiency term $\eta$. Replacing $R_{\rm XUV}$ with a (a priori unknown) multiple of the planetary radius gives the simplest form of the energy-limited approximation for atmospheric mass loss rate \citep[see, e.g.,][]{Erkaev2007,Sanz-Forcada2011}:
\begin{equation}
    \dot M=\eta \frac{3F_{\rm XUV}}{4G K \rho_p},
    \label{eq:enelim}
\end{equation}
where $G$ is the gravitational constant, $\rho_p$ is the mean planetary mass density, and the potential energy reduction factor $K<1$ accounts for the host star gravitational pull \citep{Erkaev2007}. Hereafter we shall refer to $\eta$ as evaporation efficiency (we note that this is different from the heating efficiency defined by, e.g., \citet{Salz2016b}, in that the latter does not include the unknown factor $\beta=R_{\rm XUV}/R_p$, which is embedded in our definition of $\eta$). A working definition of energy-limited thermal escape stems from adopting a fixed evaporation efficiency, typically in the range $\eta \in [0.3-1]$ \citep[see, e.g.,][]{owenwu13,odert20}. The limitations of this approximation are well documented in the literature; radiation hydrodynamics codes of varied degree of complexity all indicate that the energy-limited formula can overestimate the inferred mass outflow rates by several orders of magnitude (e.g., \citealt{krenn21} and references therein; see, however, \citealt{Cubillos2017a}). Whereas the evaporation efficiency of a single (or handful of) system can be derived by running dedicated photoionization hydrodynamics codes, a  physically-motivated prescription for $\eta$ becomes necessary for assessing the role of photoevaporation-driven mass-loss in planetary evolution studies \citep{Baraffe04,Ribas2005,Sheng14,Johnstone15,Bisikalo2018}, as well as its possible contribution to carving the observed gap \citep{fulton17} in the radius distribution of small planets \citep{owenwu13,Lammer2012,Lopez13,rogers21}. 
\begin{table}
\centering
\renewcommand{\arraystretch}{1.2}
\small
	\begin{tabular}{
    	P{1.8cm} %
    	P{1.0cm} %
    	P{0.9cm} %
    	P{1.2cm} %
    	P{1.2cm} %
    	}
    	\hline
    	\hline
                                        & %
    	\boldmath $\log\phi_p$          & %
    	\boldmath $\rho_p$                & %
    	\boldmath$\log(F_{\rm XUV})$        & %
    	\boldmath$\log(\dot{M})$  		\\
        
            & 
        (1) & 
        (2) & 
        (3) & 
        (4) \\  
    	\cline{1-5}
        55 Cnc e      & 12.44 & 6.67 & 4.37 & 10.59  \\
        GJ 1214	b     & 12.15 & 1.57 & 3.30 & 10.09  \\
        GJ 3470 b     & 12.30 & 0.82 & 3.67 & 10.65  \\
        GJ 436 b	  & 12.57 & 2.15 & 3.13 & 9.79   \\
        GJ 9827 b	  & 12.16 & 4.20 & 3.80 & 10.17  \\
        GJ 9827 d	  & 12.23 & 3.05 & 2.85 & 9.42   \\
        HAT-P-11 b    & 12.52 & 1.19 & 3.52 & 10.41  \\
        HD 149026 b   & 12.84 & 0.86 & 3.98 & 11.08  \\
        HD 189733 b   & 13.27 & 0.91 & 4.38 & 9.99   \\
        HD 209458 b   & 12.97 & 0.33 & 3.03 & 10.46  \\
        HD 97658 b	  & 12.42 & 4.78 & 2.82 & 9.24   \\
        K2-25 b       & 12.70 & 3.01 & 4.06 & 10.44  \\
        WASP-43 b     & 13.56 & 2.40 & 4.62 & 7.97   \\
        WASP-69 b	  & 12.65 & 0.27 & 4.00 & 11.50  \\
        WASP-77 A b   & 13.40 & 1.12 & 4.57 & 9.60   \\
        WASP-80 b	  & 13.02 & 0.69 & 3.77 & 10.57  \\
    	\hline
    	\hline
	\end{tabular}
	\caption{Nearby planet sample; specific gravitational potential energy (1); mean mass density (2); XUV flux at the planet orbital distance (3); mass outflow rate (4). Stellar and orbital parameters have been revised to account for the recently revised Gaia DR2 distances \citep{gaia}; X-ray fluxes were estimated from the original (Chandra or XMM-Newton) data adopting a homogeneous data reduction, and fluxes at the planet orbital distance were calculated adopting the X-ray to XUV scaling relations by \cite{king18}. The ensuing mass outflow rates are derived using ATES \citep{ates1}. From Spinelli et al. (Paper III of this series, in preparation). All quantities are expressed in cgs units.}
    \label{tab:sample}
\end{table}
\begin{figure*}
    \begin{center}
    \makebox[1.9\columnwidth]{
    \captionsetup[subfigure]{oneside,margin={1cm,0.8cm}}
     \begin{subfigure}[t]{ \columnwidth}
        \centering
	    \includegraphics[width=1 \columnwidth]{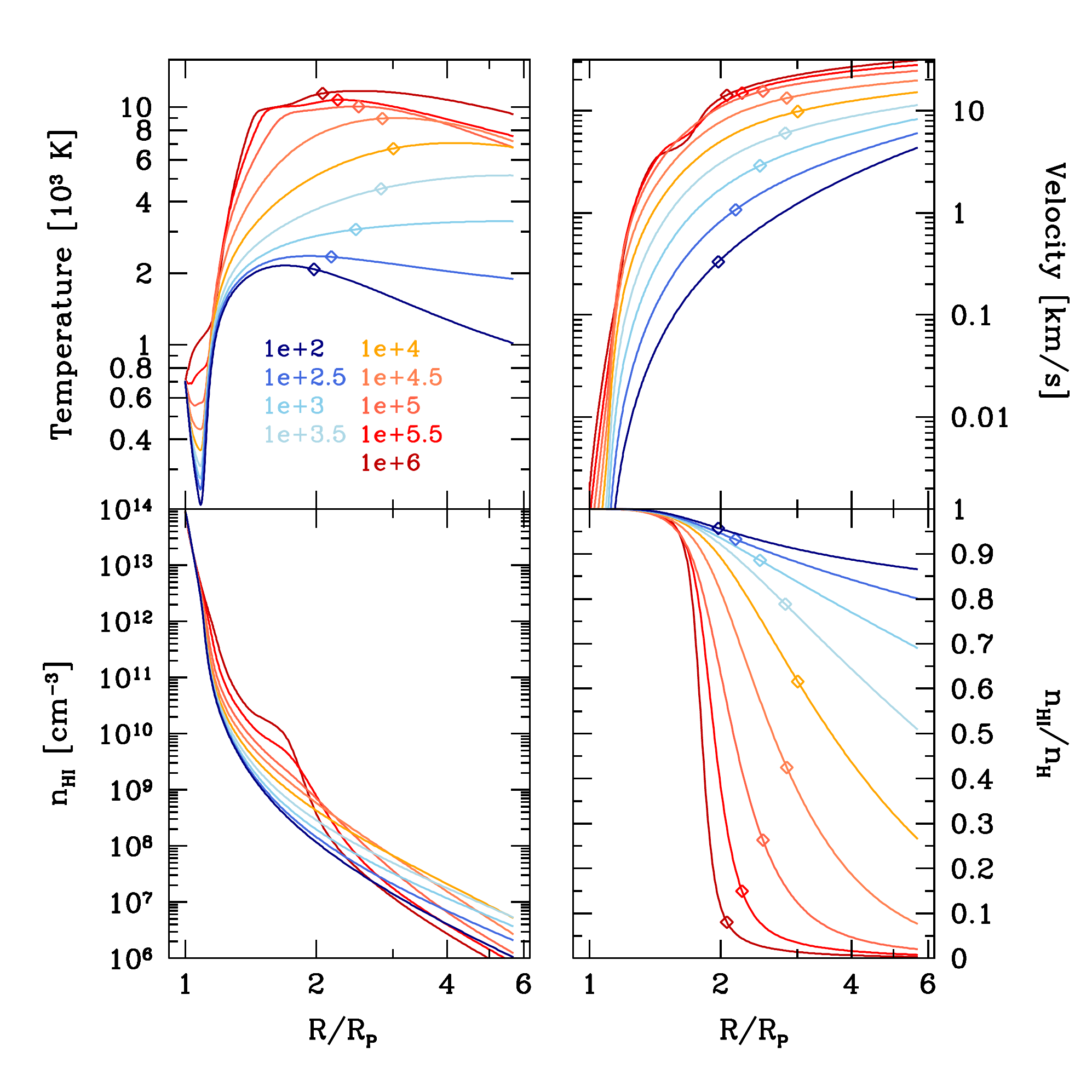}
		\caption{}
		\label{fig:low}		
    \end{subfigure}
    \captionsetup[subfigure]{oneside,margin={1cm,0.8cm}}
        \begin{subfigure}[t]{ \columnwidth}
         \centering
    	\includegraphics[width=1 \columnwidth]{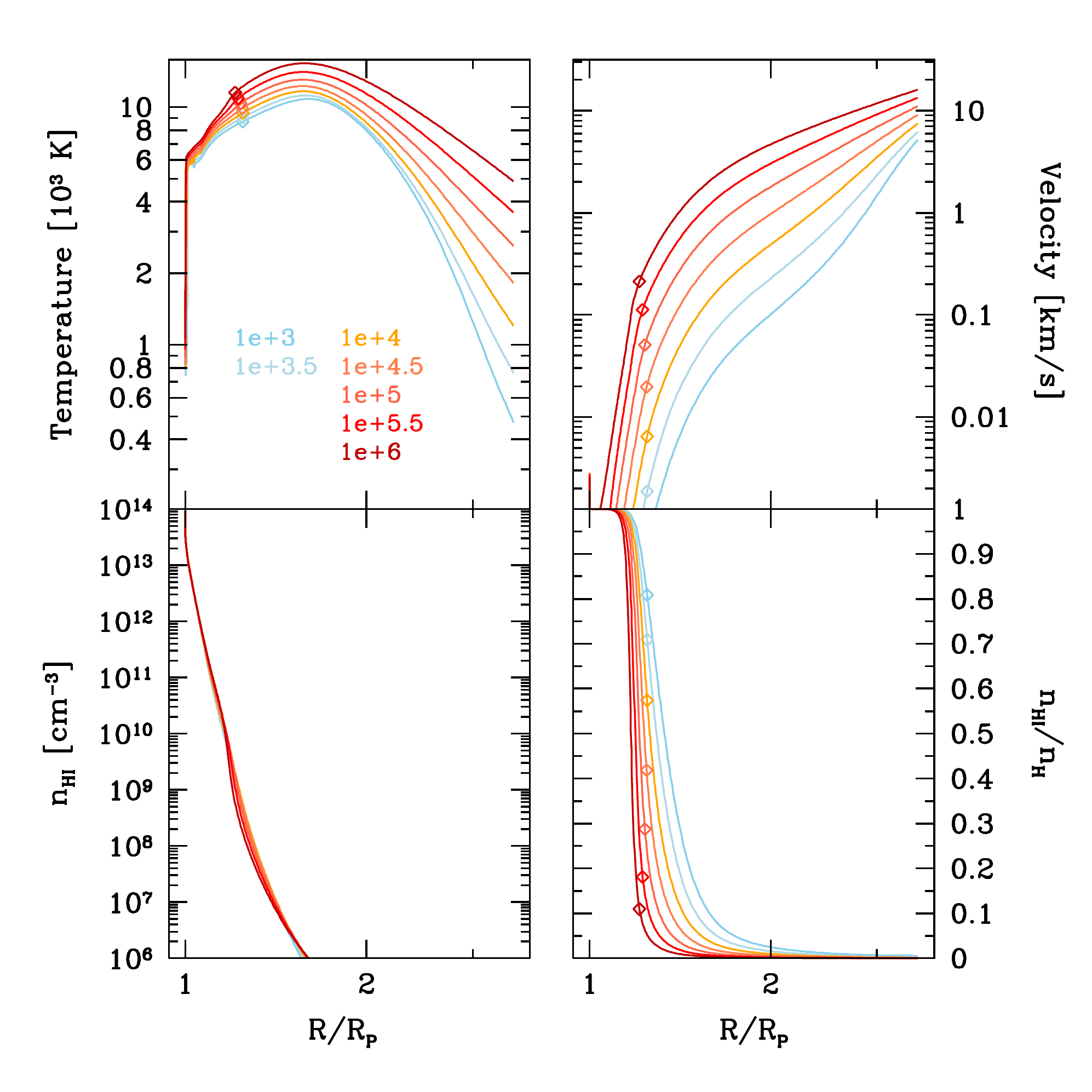}
    	\caption{}
    	\label{fig:high}
    \end{subfigure}
}
    \end{center}
    \caption{Simulated outflow properties for the Neptune-like planet GJ~3470~b (Panel (a)) and for the hot Jupiter WASP-77A~b (Panel (b)). $F_{\rm XUV}$ varies between $10^2$ (deepest blue) and $10^6$ erg cm$^{-2}$ s$^{-1}$ (deepest red), in intervals of 0.5 dex. The diamond symbols mark the location of the $\tau=1$ surface at 13.6 eV.}
    \captionsetup[subfigure]{oneside,margin={0cm,0cm}}
\end{figure*}
In a seminal paper, \cite{MC} (hereafter MC09) simulated the atmospheric escape from a Jovian-like planet subject to different irradiation levels, spanning over 5 dex in flux. They show that the energy-limited approximation tends to overestimate the inferred mass outflow rate above a flux threshold of about $\simeq 10^4$ erg cm$^{-2}$ s$^{-1}$. This is interpreted as due to the onset of radiative processes (primarily in the form of hydrogen collisional excitation, that is, Ly$\alpha$ emission) as the dominant cooling mode. \cite{owenalvarez} further expanded on this and cast the transition off of energy-limited escape in terms timescales, namely the recombination time becoming comparable and eventually much shorter than the flow timescale. \\
\indent Separately, \cite{Salz2016b} focused on the role of planetary gravity; they did so by running The Pluto-Cloudy Interface \citep[TPCI;][]{Salz2015} for a sample of more than 20 (known and artificial) planetary systems covering a wide range of masses and mass densities. This experiment reveals a sharp transition from energy-limited escape, which is valid up to specific gravitational potential energies as high as $\log(-\Phi)\simlt 13.1$ (in cgs units), to a regime where the evaporation efficiency declines sharply with increasing $\abs \Phi \abs$.\\
\indent It is worth noting that the decline in outflow efficiency with flux seen by MC09 is not nearly as sharp as that seen by \cite{Salz2016b} as a function of gravitational potential, suggesting that the latter plays a primary role in determining the nature of the outflow. Furthermore, at $\log(-\Phi)=12.94$, the test planet considered by MC09 is extremely close to the threshold value identified by Salz et al., suggesting that flux plays a non negligible (if secondary) role in the overall energy-balance. \\
\indent Although several other investigations conclude that both gravity and stellar irradiation ought to have an effect on the overall process \citep[e.g.,][among the most recent ones]{Wang2018,kuby20,kuby21,krenn21,Lampon2021}, we still lack a coherent physical understanding of the limits of energy-limited escape. This is indeed the aim of this Paper. First, we introduce our sample and data (\S~\ref{sec:sample}). Next, we focus on the effect of planetary gravity, and demonstrate analytically the existence of a threshold value above which energy-limited escape cannot take place--irrespective of the stellar irradiation level (\S~\ref{sec:phi}). \\
\indent Last, we carry out a systematic investigation of the dependence of the mass outflow rate on stellar flux over a broad range of fluxes and elucidate how the different trends exhibited by low and high-gravity planets can be broadly understood in terms of the different fractional contribution of the radiative cooling mechanisms at play (\S~\ref{sec:slopes}). We end with a summary of our main results and conclusions (\S~\ref{sec:summary}). An analytical approximation of the best-fitting evaporation efficiency is given in the Appendix. 

\begin{figure}
    \center
	\includegraphics[width=1 \columnwidth]{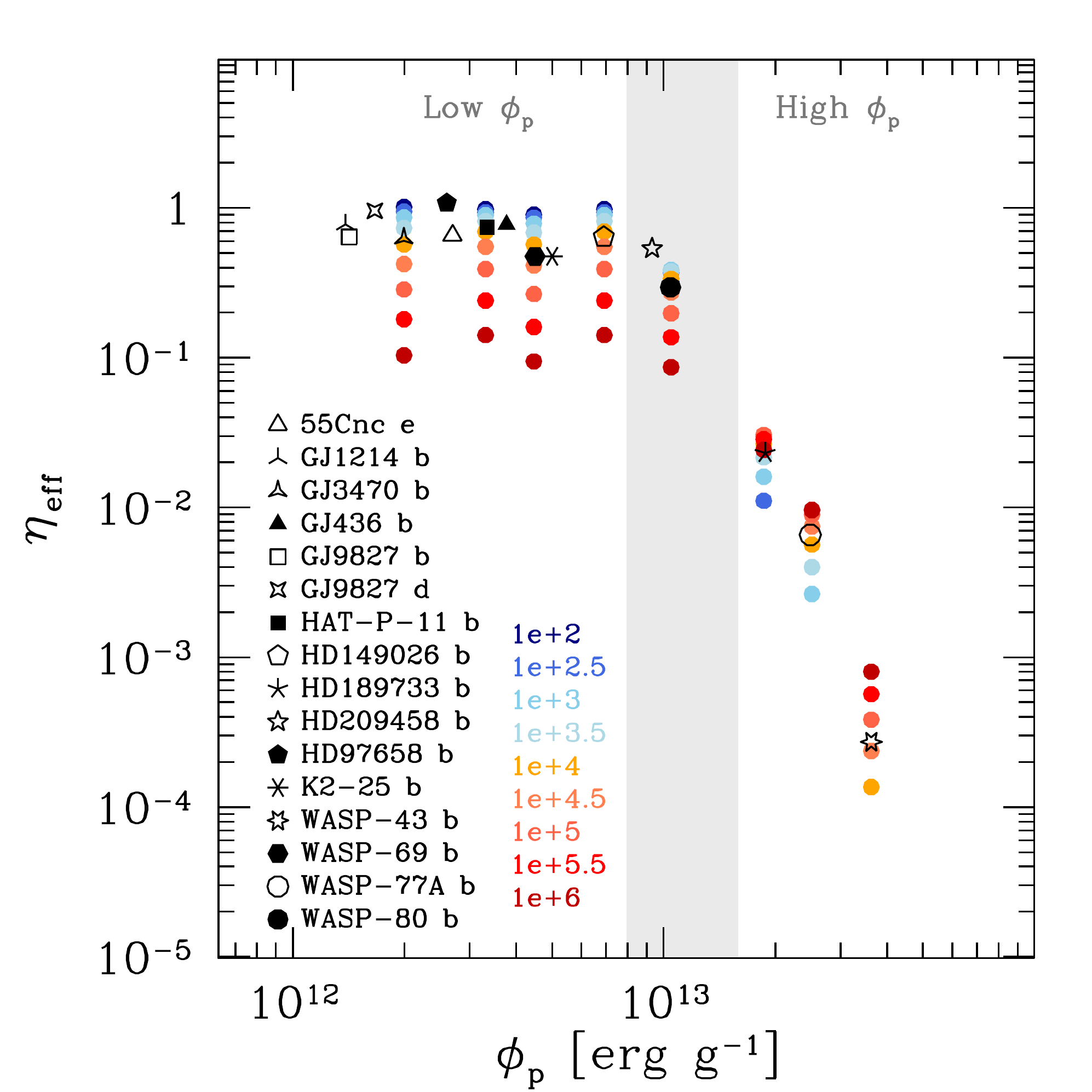}
	\caption{Outflow effective efficiency is plotted against the planet specific gravitational potential for the 16 planets considered in this work (see Table~\ref{tab:sample} for a list); $\eta_{\rm eff}$ is defined as the ratio between the ATES-estimated mass outflow rate and the energy-limited value given by Equation~\ref{eq:enelim} for $\eta=1$. To explore the effects of varied irradiation levels on $\dot{M}$, (81) simulations were carried out for a broader range of $F_{\rm XUV}$ for a subsample of eight targets. The results are represented by the colored circles, where the adopted color scheme is the same as in Figures~\ref{fig:low} and \ref{fig:high}.}
	\label{fig:etaeff}
\end{figure}
\section{Planet sample and simulated data}
\label{sec:sample}
We draw from a sample of 25 nearby ($\simlt$100 pc), moderately to highly irradiated gaseous planets, from sub-Neptunes\footnote{Here defined as having masses and radii and in the range $1.9 \leq M_p/M_\oplus \leq 10$ and $R_p/R_\oplus \geq 1.75$, respectively. We note, however, that according to the definition proposed by \citet{fulton17}, our sample would include 2 super-Earths.} through hot Jupiters, examined by Spinelli et al. (in preparation) as part of Paper III of this series. Compared to this parent sample, we limit our investigation to 16 systems with (i) $R_p>0.15 R_J$ (where the subscript $J$ indicates Jovian units), to justify the assumption of a large ($\simgt$1\%) hydrogen-helium envelope \citep{lopez14}; (ii) X-ray detected host stars; and (iii) irradiation and specific gravitational potential values within the converge range of ATES (ATmospheric EScape; a publicly available photo-ionization hydrodynamics code that was presented in Paper I of this series; \citealt{ates1}).\\
For each target Spinelli et al. present revised stellar and orbital parameters based on Gaia DR2 parallactic distances \citep{gaia}, perform a re-analysis of all the archival X-ray (Chandra and XMM-Newton) data for the purpose of uniformly estimating $F_{\rm XUV}$ (using the scaling relations of \citealt{king18}), and make use of ATES to compute the outflow temperature, ionization and density profiles, and steady-state mass loss rate. The reader is referred to Table~\ref{tab:sample} for a list of the parameters that are most relevant for this work (see tables 1 and 3 in Spinelli et al. for a complete list of parameters).   

Our targets span 1.4 dex in specific gravitational potential energy $\phi_p\equiv GM_p/R_p$ (hereafter defined as positive) vs. only 0.8 dex in $F_{\rm XUV}$. To fully parse the combined effects of planetary gravity and stellar irradiation we simulated a broader range of $F_{\rm XUV}$ for a subsample of 8 target planets, namely GJ~3470~b, HAT-P-11~b, WASP-69~b, HD~149026~b, WASP-80~b, HD~189733~b, WASP-77A~b and WASP-43~b (listed in order of increasing $\phi_p$). For each of those we aimed to run ATES adopting $F_{\rm XUV}$ values spanning between $10^2$ and $10^6$ erg cm$^{-2}$ s$^{-1}$, in intervals of 0.5 dex, for a total of 9 simulated fluxes per planet. In practice, whereas ATES can be run successfully across the whole flux range for GJ~3470~b, HAT-P-11~b, WASP-69~b, HD~149026~b and WASP-80~b the code fails to converge for the lowest-flux values in the case of the remaining (higher-gravity) planets, since the order of magnitude of the velocity acquired by the gas near the planet surface is comparable to the order of approximation of the numerical scheme. This limitation is fully consistent with the code convergence criterion laid out in Paper I of this series, namely for systems with $\log\phi_p \simgt 12.9 + 0.17\log F_{\rm XUV}$ (where the planet gravitational potential and stellar flux are expressed in cgs units).
As a result, the lowest simulated $F_{\rm XUV}$ value is $10^4$, $10^3$ and $10^{2.5}$ erg s$^{-1}$, respectively for WASP-43~b, WASP-77A~b and HD~189733~b, for a total of 81 simulations. Figures~\ref{fig:low} and \ref{fig:high} illustrate the simulation results for the case of a prototypical Neptune-like planet (GJ~3470~b, for which the simulations span over 4 dex in flux) and hot-Jupiter (WASP-77A~b, 3 dex in flux), respectively. Whereas the latter type exhibits a sharp ionization front and small range of temperatures across the flux range, the ionization and temperature profiles of the former are highly sensitive to the irradiation level. At the highest irradiation level, both reach the typical temperature of a fully-ionized gas in photo-ionization equilibrium ($\simeq 10^4$ K). \\
\indent ATES' results--both those arising from the actual sample listed in Table~\ref{tab:sample}, as well as those ensuing from the simulated $F_{\rm XUV}$ values--constitute the primary data set for our investigation below. Throughout, we illustrate the former using black symbols, and the latter using colors, with bluer (resp. redder) colors corresponding to lower (higher) irradiation, following the same color scheme as laid out in Figures~\ref{fig:low} and ~\ref{fig:high}. 

\begin{figure*}
    \captionsetup[subfigure]{justification=justified,margin={1.5cm,0.5cm}}
    \centering
    \makebox[2.0\columnwidth]{
     \begin{subfigure}[t]{\columnwidth}
        \centering
	    \includegraphics[width=0.9\columnwidth]{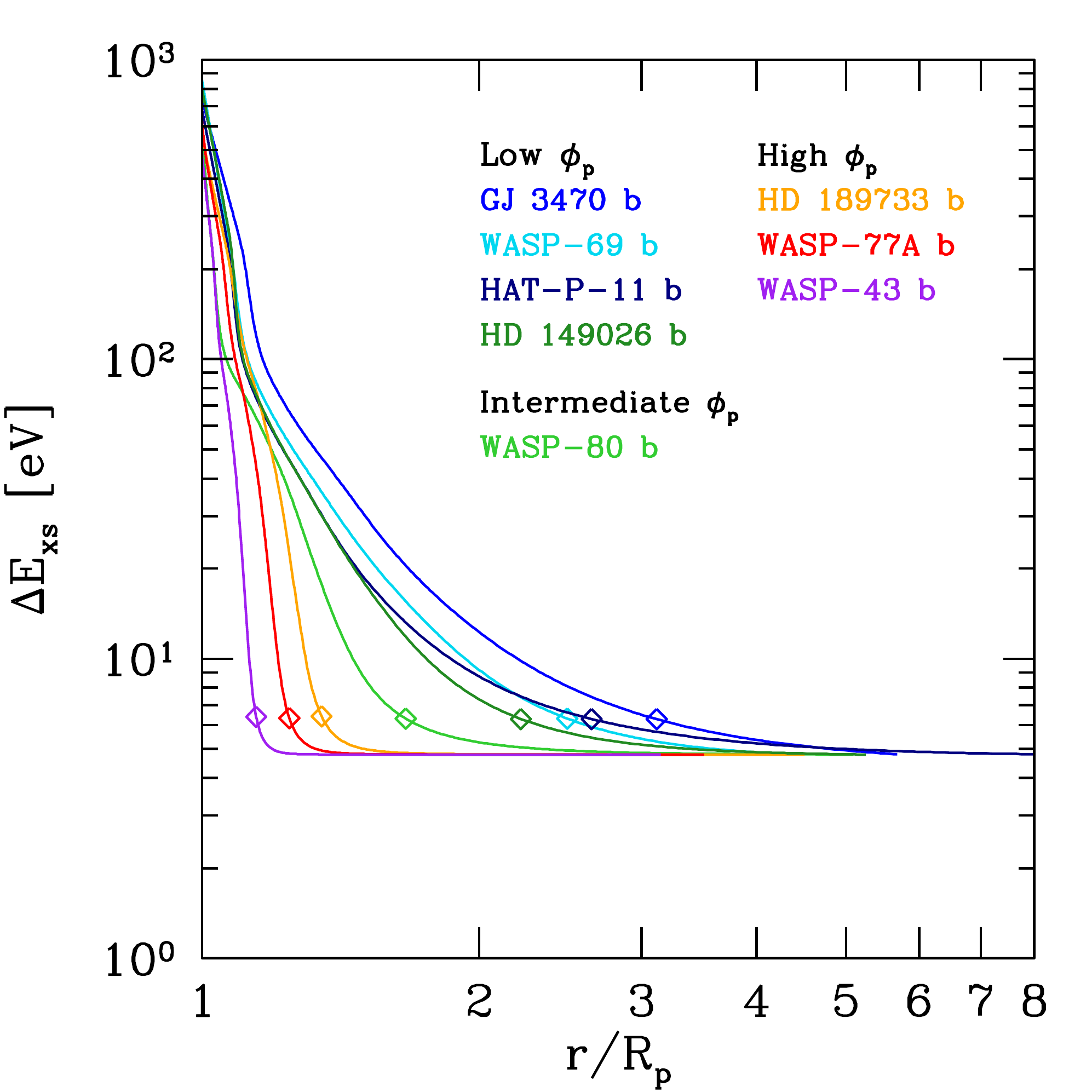}
        \caption{}
		\label{fig:Exs}		
    \end{subfigure}\hfill
    \begin{subfigure}[t]{\columnwidth}
        \centering
    	\includegraphics[width=0.9\columnwidth]{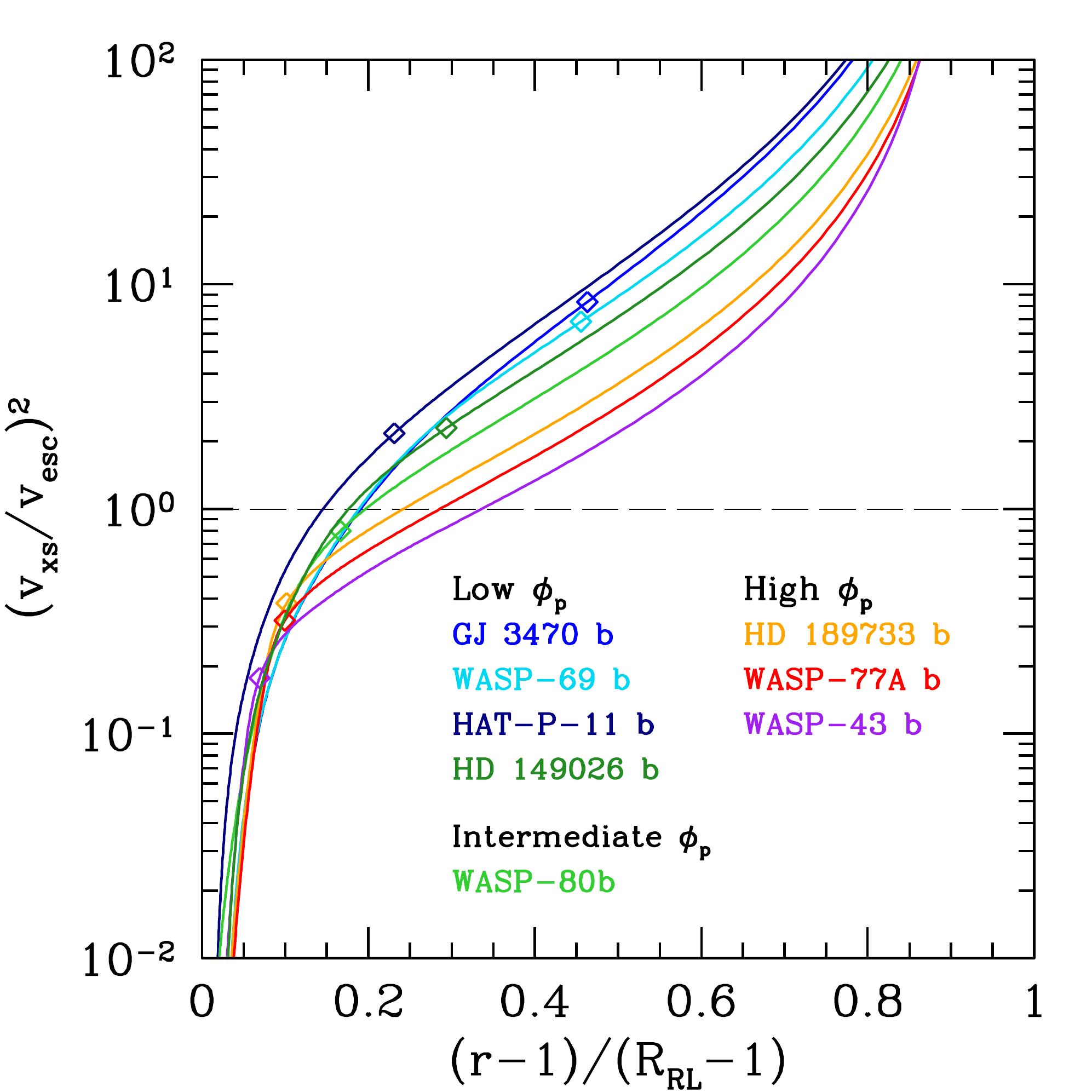}
    	\caption{}
	    \label{fig:Vxs}
    \end{subfigure}\hfill 
    }
    \caption{\HI~mean excess energy defined in Equation~\ref{eq:XS} (Panel (a)) as a function of planetary distance for different systems and Second power of the ratio between the excess and escape velocity defined in Equations~\ref{eq:vxs} and \ref{eq:vesc} (Panel (b)); $r$ and $R_{\rm RL}$ are expressed in units of the planet radius, such that the planetary distance on the $x$ axis is given as a fraction of the Roche lobe radius $R_{\rm RL}$. The dashed line represents the threshold for wind launching, i.e., where $(v_{\rm xs}/v_{\rm esc})^2=1$. In both panels, the diamond symbols indicate the location of the $\tau=1$ surface at 13.6 eV.}
    \captionsetup[subfigure]{oneside,margin={0cm,0cm}}
\end{figure*}

\section{The role of planetary gravity: $\phi_p$-limited thermal escape} 
\label{sec:phi}
Following \cite{Salz2016b}, in Figure~\ref{fig:etaeff} we plot the outflow ``effective efficiency'', $\eta_{\rm eff}$, defined as the ratio between the mass outflow rate estimated by ATES and the energy-limited expression given by Equation~\ref{eq:enelim} for $\eta=1$, as a function of $\phi_p$. This quantity enables a swift estimate of how far off from the energy-limited expression a given system is. To the extent that ATES correctly models all of the relevant physics, it also yields a quantitative estimate of the actual energy-limited evaporation efficiency.

This exercise confirms the existence of a limiting $\phi_p$ value above which the effective efficiency begins to dive well below theoretical energy-limited value.  Our results are qualitatively and quantitatively consistent with the conclusions drawn by \cite{Salz2016b}, who interpret this behavior in terms of a transition from thermally-driven outflows to a regime where radiative cooling becomes comparable to (and eventually dominates over) adiabatic cooling, thus making the outflow progressively more and more inefficient compared to the energy-limited approximation. 
For $ \log\phi_p \lesssim 13$ the bulk of the absorbed stellar flux is mostly converted into expansion work. Working against a weak gravitational pull, this yields efficient atmospheric escape. In contrast, higher gravitational potential planets host more tightly bound atmospheres and a dominant fraction of the absorbed stellar flux is re-emitted locally, primarily through Ly$\alpha$ and bremsstrahlung emission.

This widely accepted scenario, while phenomenologically correct, falls short of a quantitative explanation for the existence of such a sharp threshold, as opposed to a more gentle decline in efficiency. The heretofore unexplained  numerical value of the threshold is also of significant interest. In what follows we show analytically that the existence of a limiting $\phi_p$ at $\simeq 10^{13}$ erg g$^{-1}$ can be understood in terms of a quantitative balance between the atmospheric particles' binding energy, the photo-ionization energy budget, and the remaining energy reservoir that is available to initiate and sustain the expansion. 

Let us consider the effects of an external photo-ionizing source (that is, the stellar radiation field) on an thin layer of atmospheric gas of a planet with specific gravitational potential energy $\phi_p$. Borrowing from cosmological re-ionization theory \citep[see, e.g.,][]{Puchwein2019}, the mean excess energy $\Delta E_{\rm xs}$ within this layer can be defined as the ratio between the photo-heating rate (erg sec$^{-1}$) and the photo-ionization rate (sec$^{-1}$) per ion:
\begin{equation}
    \Delta E_{\rm xs} =\frac{\int_{\nu_t}^{\infty} {d\nu\, \frac{F_\nu}{h\nu}\sigma_\nu\, h(\nu-\nu_t})}
    {\int_{\nu_t}^{\infty} {d\nu\, \frac{F_\nu}{h\nu}\, \sigma_\nu}},
\label{eq:XS}
\end{equation}
where $F_\nu=F_{\nu,0}\exp(-\tau_\nu)$ is the local, absorbed stellar flux at frequency $\nu$, $\sigma_\nu$ is the photo-ionization cross section, $\tau_\nu$ is the optical depth, $\nu_t$ is the photo-ionization frequency threshold and $h$ is the Planck constant.
In practical terms, this quantity represents the amount of energy that is still available to heat up the gas after a single photo-ionization event \citep[see, e.g.,][]{Osterbrock2006}. 
We postulate that, for an atmospheric particle of mass $m$ to be able to escape the system under the effect of photo-heating, the volume-averaged mean excess energy must exceed the particle binding energy: $\langle \Delta E_{\rm xs} \rangle \simgt \phi_p m$.

The crux of this approach is that $\Delta E_{\rm xs}$ is independent of the normalization of the photo-ionizing flux, as it only depends on the shape of the radiation spectrum close to the ionization threshold of the element(s) under consideration.
For simplicity, we start by considering the case of a pure hydrogen atmosphere. To first order the shape of the photo-ionization cross-section close 13.6 eV scales with the frequency as $\nu^{-2.75}$, such that, given an external, power-law shaped photo-ionizing radiation spectrum $\propto \nu^{-\beta}$, the mean excess energy can be approximated as $\Delta E_{\rm xs}\simeq 13.6/(\beta+1.75)$ eV. In the optically-thin regime, such as for a typical unabsorbed stellar spectrum with $\beta\simeq 1$, $\Delta E_{\rm xs}\simeq 5$ eV. However, owing to the effect of atmospheric absorption, the spectrum becomes progressively harder deeper into the atmosphere. In fact, $F_{\nu}$ becomes nearly flat above 13.6 eV just across the ionization front (see, e.g., figure 5 in \citealt{Haardt1996} for a cosmological application), implying that $\Delta E_{\rm xs} \simeq 8$ eV in the layer where the bulk of stellar radiation is absorbed. 

For a more robust, quantitative estimate we used ATES to simulate the atmospheres of eight case studies, spanning 1.4 dex in $\phi_p$. As illustrated in Figure~\ref{fig:Exs}, $\Delta E_{\rm xs}$ varies within a remarkably narrow range (between $5-7$ eV) within the optically thin portion of the outflow (defined by $\tau \leq 1$ at 13.6 eV). Only the most energetic (and thus rare) photons penetrate deeper, into the optically thick portion, producing (rare) ionization events with extremely high values of $\Delta E_{\rm xs}$. Using the atmospheric density and ionization profiles obtained by ATES for the same eight planets, also under different irradiation regimes (for a total of 81 simulations, spanning 4 dex in $F_{\rm XUV}$) we calculated the volume-averaged mean excess energy by integrating $\Delta E_{\rm xs}$ over the free-electron number density. This yields values in the range $\langle \Delta E_{\rm xs} \rangle \in [8-15]$ eV. 

The minimal condition for the onset of a thermally-driven outflow can thus be recast in terms of an upper limit to the specific gravitational potential. By setting $m$ equal to the proton rest mass $m_p$ we obtain: { $\phi_p \simlt \phi_p^{\rm thr} \approx [7.7-14.4]\times 10^{12}$ erg g$^{-1}$}, or~ $\log\phi_p \simlt \log\phi_p^{\rm thr} \approx [12.9-13.2]$ (in cgs units). 

Several effects complicate the matter, such as the presence of heavier elements than hydrogen (primarily helium), the possible presence of hydrogen molecules in the lower atmosphere, and the fact that the outflow is launched past $R_p$, implying a slightly reduced binding energy. Each is expected to carry a correction factor of order a few to the above criterion. Overall, we find a remarkably good agreement with the empirical value of $\sim 10^{13}$ erg g$^{-1}$ found by \cite{Salz2016b} and subsequent work (including the estimates presented in Figure \ref{fig:etaeff}), all of which make use of radiation hydrodynamics codes that properly model the wind launching mechanism and also account for the presence of helium. 
Hereafter, we shall loosely refer to planets with $\log\phi_p<12.9$ and $\log\phi_p>13.2$ as low- and high-gravity planets, respectively.
\begin{figure*} 
    \captionsetup[subfigure]{justification=justified,margin={1.5cm,0.5cm}}
    \begin{center}
    \makebox[1.9\columnwidth]{
     \begin{subfigure}[t]{ \columnwidth}
         \centering
         \includegraphics[width = 1.0\columnwidth, keepaspectratio]{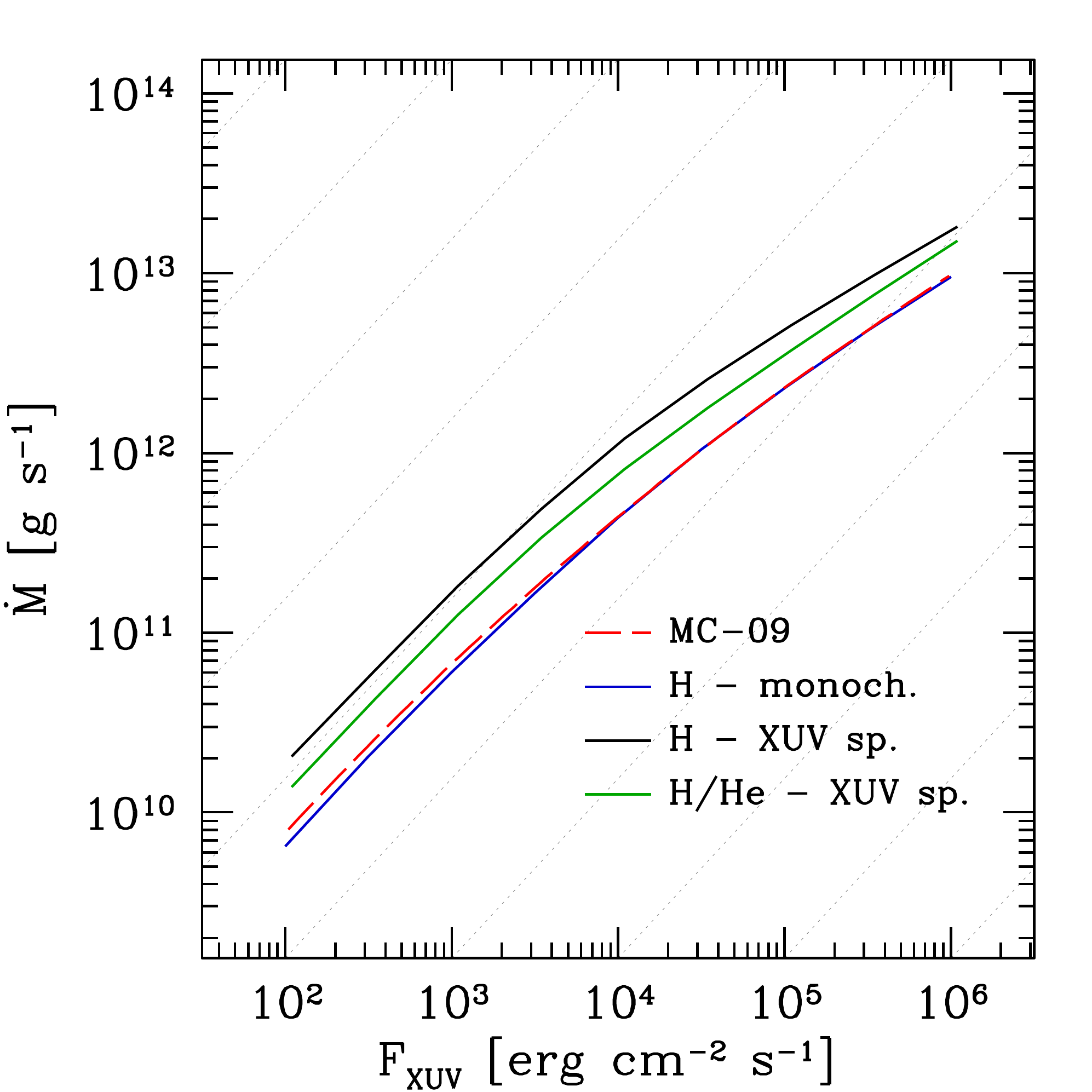}
	\caption{}\label{fig:mc}
     \end{subfigure}\hfill
     \captionsetup[subfigure]{justification=justified,margin={0.75cm,0.25cm}}
     \begin{subfigure}[t]{ \columnwidth}
         \centering
         \includegraphics[width = 1.0\columnwidth, keepaspectratio]{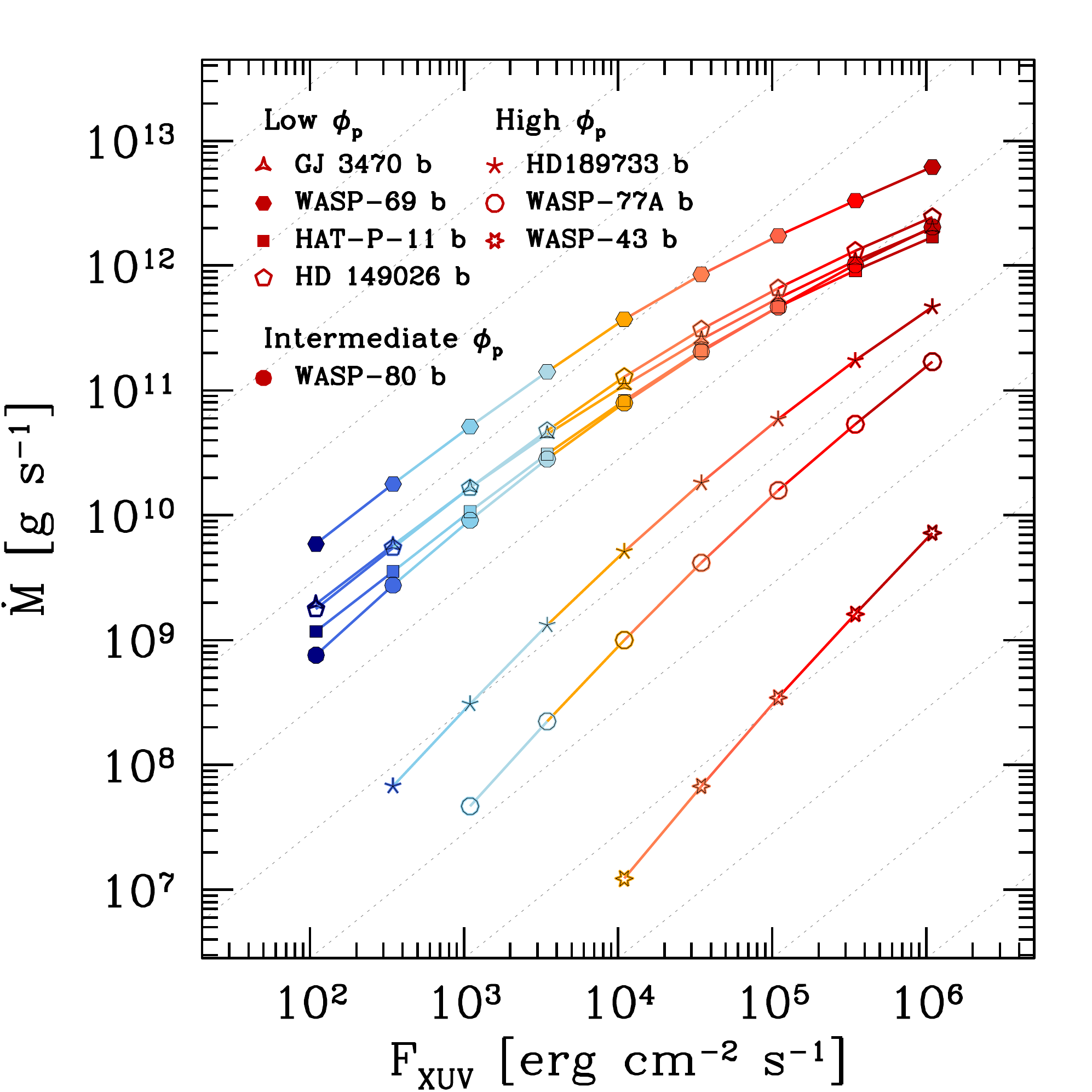}
         \caption{}\label{fig:MCexp_FvsMdot} 
     \end{subfigure}\hfill
    }
    \end{center}
    \caption{Simulated mass loss rate $\dot{M}$ for an artificial planet with $M_p=1.3\times 10^{30}$ g and $R_p=10^{10}$ cm, subject to different levels of stellar XUV irradiation (Panel (a)) and simulated mass outflow rates for eight planets subject to different levels of stellar XUV irradiation, for a total of 81 simulations (Panel (b)). In Panel (a), the dashed red line represents the results of the simulations carried out by \cite{MC}, under the assumption of a pure hydrogen atmosphere and mono-energetic irradiation field at 20~eV. The blue line represents the results obtained by ATES under the same assumptions. The green line represents the ATES results for a pure hydrogen atmosphere subject to a power-law irradiation spectrum with slope $\nu^{-1}$. The black line refers to a primordial hydrogen-helium atmosphere with a power-law spectrum, that is, the standard setup for ATES. The adopted color scheme in Panel (b) is the same as shown in Figures~\ref{fig:low}, \ref{fig:high} and \ref{fig:etaeff}. In both panels the dotted lines represent linear proportionality between mass loss rate and flux.}
\end{figure*}
%

Introducing the concept of mean excess energy enables a novel definition of the outflow launching point, as follows. 
At each radius, $\Delta E_{\rm xs}$ represents the mean energy carried by a photo-electron;
this energy is thermalized locally through collisions, such that the mean kinetic energy gained by a proton in the process can be written as: 
\begin{equation}
     m_p\frac{v_{\rm xs}^2}{2} \equiv \frac{n_e}{n_\Hy}\Delta E_{\rm xs}.
    \label{eq:vxs}
\end{equation}
Here $n_\Hy$ and $n_e$ are the hydrogen and free electron number density, respectively, and the ``excess velocity'' $v_{\rm xs}$ is the mean velocity that is acquired locally by a proton. The outflow launching point is identified by the radius (as in distance from the planet center) at which the excess velocity exceeds the escape velocity: $(v_{\rm xs}/v_{\rm esc})^2 \simgt 1$, where $v_{\rm esc}$ is calculated through the following expression:
\begin{equation}
    \frac{v_{\rm esc}^2(r)}{2} = \left[\Phi(R_{\rm RL})- \Phi(r)\right].
    \label{eq:vesc}
\end{equation}
$\Phi(r)$ is the gravitational potential of the planet-star system, defined in equation 2 of \cite{ates1}, $R_{\rm RL}$ refers to the Roche Lobe radius, while $r$ is the distance from the planet center. 

The above arguments can be easily generalized for a mixed hydrogen-helium atmosphere. In this case, Equation~\ref{eq:vxs} becomes: 
\begin{equation}
 \mu \frac{v_{\rm xs}^2}{2} \equiv  f_\HII \, \Delta E_\HI +Y(f_\HeII \, \Delta E_\HeI + 2 f_\HeIII \, \Delta E_\HeII),
\label{eq:vescHe}
\end{equation}
where \HI\ and \HII\ denote neutral and ionized hydrogen, while \HeI, \HeII\ and \HeIII\ denote neutral, single-ionized and double-ionized helium; $f_i$ and $\Delta E_i$ are the ionization fraction and mean excess energy for the relevant $i$ species, $Y\equiv n_\He/n_\Hy$ is the helium to hydrogen number abundance ratio and $\mu\equiv (1+4Y)m_p$. 

Figure~\ref{fig:Vxs} compares the location of the outflow launching point as identified by the $(v_{\rm xs}/v_{\rm esc})^2 \simgt 1$ criterion with the location of the $\tau=1$ surface (calculated at 13.6 eV throughout), for the same eight planets shown in Figure~\ref{fig:Exs}. Interestingly, our newly defined outflow launching point approximately coincides with the $\tau=1$ surface only for the intermediate-gravity planet WASP-80~b, whereas it is located within (outside) the $\tau=1$ surface for low- (high-) gravity planets. This has profound consequences for the outflow evaporation efficiency, since the $\tau = 1$ surface sets the location where the stellar energy absorption starts to become significant (in fact, the $\tau=1$ surface is often adopted as a working definition for the outflow launching point in the literature).
For low-gravity planets, the shallow ionization profiles (see, e.g., Figure~\ref{fig:low}) guarantee that a sizable fraction of the absorption occurs within an extended layer located between the $\tau\simgt 1$ and the $v_{\rm xs}\simeq v_{\rm esc}$ radius. This means that most of the energy is disposed of where (on average) the ions have sufficiently high kinetic energy to escape the system. As illustrated in the next Section, this regime corresponds to evaporation efficiencies in the range $\in$[0.1-1], depending on the irradiation level.

In contrast, the sharp ionization fronts of high-gravity planets (see, e.g., Figure~\ref{fig:high}) imply that only a small fraction of the stellar flux is absorbed above the $(v_{\rm xs}/v_{\rm esc})^2= 1$ line, where the ionization fractions are high and the densities are low. Most of the energy absorption occurs within a very narrow layer where (after thermalization) the average ion kinetic energy is insufficient to warrant escape. Combined, these effects are responsible for the sharp decline in evaporation efficiency above $\phi_p$, down to values as low as $10^{-4}$.


\section{Flux-dependent evaporation efficiency}\label{sec:slopes}

We turn our attention to the functional dependence of the outflow rate on irradiation. 
MC09 showed that for an artificial planet with $M_p= 1.3\times 10^{30}$ g and $R_P=10^{10}$ cm (or $\log\phi_p=12.94$) the transition from energy-limited thermal escape to a regime where radiative cooling starts to dominate occurs close to $\simeq 10^4$ erg cm$^{-2}$ s$^{-1}$. To start with, we verify that ATES reproduces the same results as found by MC09. To do so, we modify the code to simulate a monochromatic radiation field and pure hydrogen atmosphere. The results of this comparison are illustrated in Figure~\ref{fig:mc}. While relaxing those approximations results in higher $\dot{M}$ across the flux range, we recover the same behavior as described MC09, namely that $\dot{M}$ scales linearly with $F_{\rm XUV}$ below $\simlt 10^4$ erg cm$^{-2}$ s$^{-1}$. The slope flattens above it, signaling a departure off of energy-limited thermal escape.\\
\indent MC09 attributed the change is slope in the $\dot{M}$:$F_{\rm XUV}$ relation seen around $F_{\rm XUV} \simeq 10^4$ erg cm$^{-2}$ s$^{-1}$ to the transition from an energy-limited regime to one where radiative cooling, specifically in the form of Ly$\alpha$, dominates the flow. This argument was expanded upon by \cite{owenalvarez}, who differentiate between an energy-limited regime and a recombination-limited regime. In the former the recombination time is much longer than the flow timescale, the outflow exhibits a very extended ionization front, and the mass outflow rate can be expected to scale linearly with the ionizing flux. This regime is typical of low-gravity planets at low irradiation. At high irradiation levels the ionization front becomes sharp, the flow is in a radiative-recombination equilibrium state, and the mass-loss rate scales approximately as the square root of the ionizing flux. \cite{owenalvarez} also show that the transition is rather slow, in the sense that it occurs over a large range of fluxes, with higher gravity planets transitioning to the recombination-limited regime at lower fluxes (we note that the highest-gravity planet simulated by \cite{owenalvarez} has $\log\phi_p=12.6$, and is thus in the low-gravity regime as per \S~\ref{sec:phi}). 

\begin{figure} 
         \centering
         \includegraphics[width = 1.0\columnwidth, keepaspectratio]{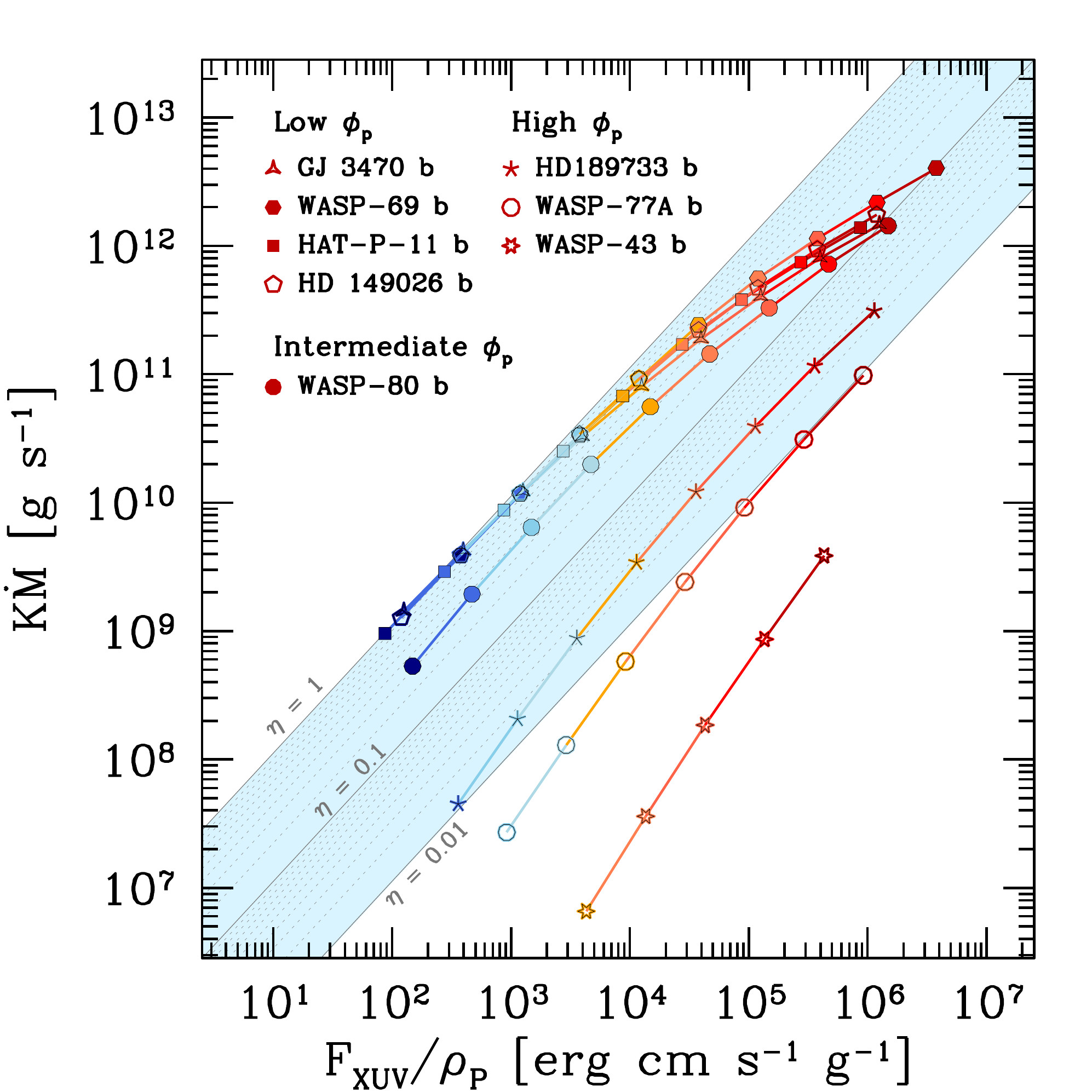}
         \caption{Simulated, $K$-reduced mass outflow rate, shown as a function of the irradiation to planet mass density ratio, for the same planets shown in Figure~\ref{fig:MCexp_FvsMdot}. Energy-limited escape yields $K \dot{M}=F_{\rm XUV}/\rho_p$ (i.e., $\eta=1$ in Equation~\ref{eq:enelim}). The adopted color scheme is the same as shown in Figures~\ref{fig:low}, \ref{fig:high} and \ref{fig:etaeff}. }
         \label{fig:MCexp_FrhovsKMdot} 
\end{figure}

In order to further investigate the relation between $\dot{M}$ and $F_{\rm XUV}$ across the full range of $\phi_p$ and $F_{\rm XUV}$ we rely on the simulations described in \S\ref{sec:sample}. As a reminder, these include four low-gravity planets (GJ~3470~b, WASP-69~b, HAT-P-11~b and HD~149026~b), three high-gravity planets (HD~189733~b, WASP-77A~b and WASP-43~b), plus WASP-80~b, which, at $\log\phi_p=13.02$ can be considered as an intermediate case. All of those were simulated over a wide range of $F_{\rm XUV}$ spanning from $10^2$ to $10^6$ erg cm$^{-2}$ s$^{-1}$, in intervals of 0.5 dex, for a total of 81 simulations. The results are illustrated in Figure~\ref{fig:MCexp_FvsMdot}.
Overall, it is interesting to note how different planets exhibit different slopes as well as normalizations--both compared to the $\dot{M}$:$F_{\rm XUV}$ relation seen for the test-planets considered by MC09 (Figure~\ref{fig:mc}) and \cite{owenalvarez}, as well as different from one another. 
This is best seen in Figure~~\ref{fig:MCexp_FrhovsKMdot}, where we plot the $K$-reduced mass outflow rate\footnote{The $K$ term (see equation 17 in \citealt{Erkaev2007}) varies between $0.5 \lesssim K \lesssim 1$ for our sample.} as a function of the stellar irradiation to planetary density ratio ($F_{\rm XUV}/\rho_p$) for the sample under consideration. 
Since, for an energy-limited outflow, the product $K \dot{M}$ is expected to scale linearly with this quantity along the $\eta= 1$ line (Equation~\ref{eq:enelim}) downward deviations readily pinpoint any departures from the energy-limited formalism. 
Our simulations confirm that energy-limited escape is attained by low-gravity planets only up to $F_{\rm XUV}/\rho_p \simlt 10^{3-3.5}$ g cm s$^{-1}$ g$^{-1}$. The curves flatten at higher fluxes, and the outflow efficiency decreases (by a factor up to 10) with increasing irradiation. 

\begin{figure*}
    \centering
    \includegraphics[width = .9\textwidth]{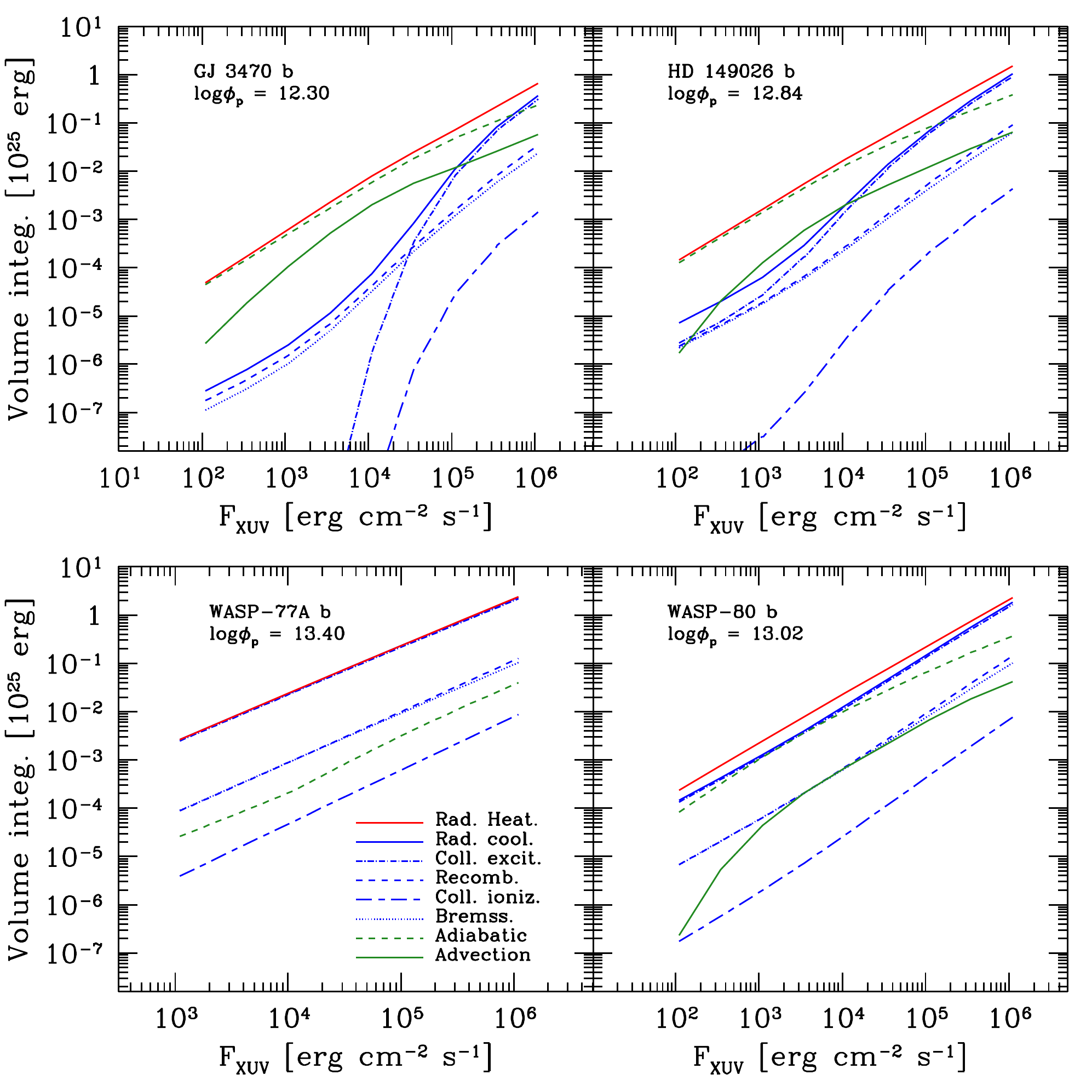}
    \caption{ Volume integrals of the various cooling processes of our model as functions of the impinging XUV flux. From top-left, moving clockwise, the four panels represent different planets with increasing $\phi_p$, from : GJ 3470 b ($\log\phi_p = 12.30$), HD 149026 b ($\log\phi_p = 12.84$), WASP-80 b ($\log\phi_p=13.02$) and WASP-77A b ($\log\phi_p = 13.40$). The various contributions have been calculated using the simulations presented in \S\ref{sec:sample}. The first two objects fall in the range of low gravity planets, the third one is an intermediate gravity planet, while the last one is a genuine high gravity planet. The advection contribution is not shown for WASP-77A b, since it is of the same order of magnitude the order of the numerical approximation of the code. }
    \label{fig:volume_integs}
\end{figure*}

In contrast, the high-gravity planets all exhibit much lower effective efficiencies ($\eta\lesssim 0.05$). However, their outflows tend to become more efficient for progressively higher irradiation values, that is the slope of the $K\dot{M}$ vs $F_{\rm XUV}/\rho_p$ curves for high-gravity planet is steeper than linear (the intermediate gravity planet WASP-80~b --solid circles in Figure~\ref{fig:MCexp_FvsMdot}-- exhibits somewhat of a hybrid behavior). 
The same effect can be visualized by looking at the colored points in Figure~\ref{fig:etaeff}, where different colors correspond to different irradiation levels (the adopted color scheme is the same as in Figures~\ref{fig:MCexp_FvsMdot} and \ref{fig:MCexp_FrhovsKMdot}). Whereas low-gravity planets show a decrease in effective efficiency with increased fluxes the opposite is true for high-gravity planets. We refer to this behavior as `evaporation efficiency inversion'.\\
\indent Next, we examine this phenomenon in terms of fractional contributions from the competing cooling processes to the overall energy budget, and specifically adiabatic, advection and radiative cooling. This is shown in Figure~\ref{fig:volume_integs}, where the volumetric contributions of each are plotted as a function of flux for four case studies.

\subsection{Low-gravity planets: Flux-limited thermal escape}
\label{sec:lowgrav_flux}

We start by discussing the behavior of GJ~3470~b as our low-gravity case study ($\log\phi_p = 12.3$).  Planets such as GJ~3470~b exhibit a wide range of atmospheric temperatures and ionization fractions as the irradiation level changes. The atmospheric temperature at the $\tau=1$ surface varies by about one order of magnitude as the flux increases from $10^{2}$ to $10^6$ erg sec$^{-1}$ cm$^{-2}$ (see Figure~\ref{fig:low}). At the lowest flux, the ionization front is very broad and virtually 100\% of the absorbed stellar radiation is converted into adiabatic expansion (top left panel of Figure~\ref{fig:volume_integs}).
Adiabatic cooling (dashed green line) starts to dive (albeit gently) below $\simeq 100\%$ already at $\simeq 10^{3}$ erg sec$^{-1}$ cm$^{-2}$. Above this flux level, the flow ceases to be formally energy-limited and its efficiency decreases with increasing flux owing to the progressive increase of (i) advective cooling (solid green line) and (ii) radiative cooling (solid blue), primarily in the form of collisional excitation (Ly$\alpha$). 
The fractional contribution of advective cooling (which scales a $\dot{M}T/\mu$) peaks at $\simeq 10^{4}$ erg sec$^{-1}$ cm$^{-2}$ (where it accounts for 26\% of the total cooling), and decreases at higher fluxes. This happens as a result of the fact that the flow temperature approaches the thermostat value of $10^4$ K, while the flow approaches full ionization. \\
\indent However, Ly$\alpha$ starts to skyrocket\footnote{This sharp rise is driven by the exponential dependence of the Ly$\alpha$ cooling coefficient on atmospheric temperature; $\Lambda_{Ly\alpha} \propto \exp(-T_0/T)$ s$^{-1}$, where $T_0$=118,348 K (the complete functional form $\Lambda_{Ly\alpha}$ is given in Appendix A of Paper I of this series; \citealt{ates1}).} above recombinations and bremsstrahlung losses right around the same flux where advective cooling reaches its peak fractional contribution. At $F_{\rm XUV}\simgt 10^5$ Ly$\alpha$ dominates over advective cooling.\\
\indent As the flux further increases the outflow efficiency decreases at a progressively greater pace (down to about 10\%) due to the nonlinear growth of the Ly$\alpha$ cooling contribution. While the actual flux above which Ly$\alpha$ losses dominate over advective cooling depends on the planet gravity, the same qualitative description applies to all low-gravity planets ($\log\phi_p\simlt 12.9$). For those, advective and radiative (Ly$\alpha$) losses, combined, drive the progressive decline of the evaporation efficiency with increasing irradiation. More specifically, as also shown by \cite{owenalvarez}, the transition toward a radiation-dominated regime occurs at lower fluxes for higher gravity planets (top right panel of Figure \ref{fig:volume_integs} for the case of HD~149026~b, with $\log\phi_p = 12.8$). 

\subsection{High-gravity planets}

As we argue in \S~\ref{sec:phi}, high-gravity planets can never sustain energy-limited outflows. As illustrated in Figure \ref{fig:high} for the case of WASP-77A~b, the outflow density is typically so low that the ionization structure resembles an inverted Str\"omgren sphere, exhibiting a very sharp transition between an outer fully ionized region and an inner, neutral region (this happens even at relatively low irradiation, contrary to the case of light planets). Irrespective of the irradiation level, the atmosphere temperature close to the ionization front approaches $(1-2) \cdot 10^4$ K, that is, the thermostat temperature of a highly ionized gas in ionization equilibrium. As a consequence Ly$\alpha$ cooling approaches a saturation level, in the sense that, unlike for the low-gravity planets, increasing the irradiation level does not result in a sensible increase of the fractional contribution in Ly$\alpha$ cooling.  This can be seen in the bottom left panel of Figure~\ref{fig:volume_integs}. Ly$\alpha$ accounts for close to 90\% of the overall cooling across the whole flux range; if anything, its fractional contribution decreases slightly, from 92 to 88\%, in going from $F_{\rm XUV}=10^3$ to $10^6$ erg cm$^{-2}$ s$^{-1}$, while the fractional contributions from recombinations and bremsstrahlung increase from 3 to 5\% and from 3 to 4\%, respectively, over the same range. The remaining energy (1-2\%) goes into adiabatic expansion; advection losses are completely negligible in this regime.\\
\indent  The case of WASP-80~b (bottom right panel of Figure \ref{fig:volume_integs}) is intermediate between low- and high-gravity planets. The relative contributions from adiabatic and radiative cooling are similar up to $F_{\rm XUV}\sim 10^4$ erg cm$^{-2}$ s$^{-1}$, above which Ly$\alpha$ losses take over. Advective losses are always subdominant. 
This causes the efficiency to increase ever so slightly with flux up to $F_{\rm XUV}\sim 10^4$ erg cm$^{-2}$ s$^{-1}$, above which it starts to decrease with increasing flux, thus resembling low-gravity planets. 

In summary, as the irradiation flux increases, the fractional contribution of adiabatic cooling to the overall process--however small--grows more rapidly than the radiative component. As a result, the evaporation efficiency--however small--tends to increase with flux in high-gravity planets. 

\section{Summary and conclusions}\label{sec:summary}
In this Paper, we make use of the ATES 1D photo-ionization hydrodynamics code \citep{ates1} to carry out a systematic assessment of the combined effects of planetary gravity and stellar irradiation upon atmospheric escape of primordial (hydrogen-helium) atmospheres, with a particular focus on the limits of validity of the so called energy-limited approximation (Equation~\ref{eq:enelim}). 
To investigate as broad as possible a parameter space, we simulated a set of 16 known nearby sub-Neptunes and hot-Jupiters (Table~\ref{tab:sample}) over a broad range of irradiation levels. The target planets have specific gravitational potential energy between $12.15 \simlt\log\phi_p\simlt 13.56$, and are selected to have radii below 0.15$R_J$, so as to reasonably fulfill the assumption of an extended hydrogen-helium atmosphere \citep{lopez14}. For each, we run ATES for input values of stellar XUV flux at the orbital distance ($F_{\rm XUV}$) varying between 3-4 dex, in intervals of 0.5 dex. Our main results can be summarized as follows. 

\begin{itemize}
\item Borrowing from cosmological re-ionization theory, we define the atmospheric mean excess energy as the ratio between the photo-heating rate and the photo-ionization rate per ion, and postulate that, for an atmospheric particle to be able to escape the system efficiently under the effect of photo-heating, the volume-averaged mean excess energy must exceed the particle binding energy. 
This simple criterion yields a (flux-independent) limiting $\phi_p^{\rm thr}$ below which energy-limited outflows may be attained: $\log\phi_p \simgt \log\phi_p^{\rm thr}\approx [12.9-13.2]$ (in erg g$^{-1}$, see Figure~\ref{fig:etaeff}). This range agrees remarkably well with the phenomenological threshold that was previously identified in the literature \citep{Salz2016b}, and cements the notion that the energy-limited formula (that is, Equation~\ref{eq:enelim} with $\eta\simeq$ 0.3-1) severely overestimates the mass outflow rates for gas giants.  

\item Whether planets with $\phi_p\simlt \phi_p^{\rm thr}$ can be expected to drive energy-limited outflows mainly depends on the XUV irradiation level. As also shown by previous work \citep{owenalvarez}, the transition from energy-limited to a regime where cooling is dominated by Ly$\alpha$ is rather slow and occurs over a broad range of fluxes ($\flux \simeq 10^{3.5-5}$~erg~cm$^{-2}$s$^{-1}$), with higher gravity planets becoming radiation-dominated at lower fluxes. 

\item The introduction of an atmospheric mean excess energy translates into a novel definition of the outflow launching site, that is, the location where the mean excess velocity acquired by an ion by thermalization of a photo-electron exceeds the local escape velocity (Figure~\ref{fig:Vxs}).  Crucially, most of the stellar energy absorption occurs above this height in the case of low gravity planets, implying that the evaporation efficiency is largely independent of $\phi_p$ for those planets. Conversely, in the case of high-gravity planets only a small fraction of the energy is absorbed at heights where the ion excess velocity is sufficient high to warrant escape. As a result, the overall evaporation efficiency declines sharply as $\phi_p$ increases. 

\item 
To further elucidate the dependence of evaporation efficiency (defined in Equation~\ref{eq:enelim}) on irradiation we examine the relative contribution of adiabatic vs. advective and radiative cooling to the overall energy budget (Figure~\ref{fig:volume_integs}). Energy-limited escape ($\simeq$100\% efficiency) only occurs for planets below $\phi^{\rm thr}_p$ at the lowest irradiation, that is when advective and radiative losses are negligible. As $\flux$ increases above $\simgt 10^{3-3.5}$~erg~cm$^{-2}$s$^{-1}$ the evaporation efficiency starts to decline, by up to an order of magnitude. This trend stems from a progressive increase in the relative contribution of advective cooling, followed by a sharp surge in Ly$\alpha$ losses (Figure~\ref{fig:volume_integs}).

\item 
In spite of the much lower absolute values (a few per cent or lower) compared with lower gravity planets, the evaporation efficiency of planets above $\phi^{\rm thr}_p$ increases with irradiation (Figure~\ref{fig:MCexp_FrhovsKMdot}). These systems are characterized by sharp ionization fronts and (nearly) fully ionized outflows, which implies a saturation of the Ly$\alpha$ cooling channel. As a result, the fractional contribution of adiabatic cooling (albeit low) increases with flux at a faster pace than Ly$\alpha$.


\end{itemize}

In closing, these results allow for the rapid, physically motivated characterization of the expected atmospheric evaporation efficiency $\eta$ for a given planet, thus enabling the community to move past the ``one-size-fits-all'' approach. An analytical approximation of the best-fitting effective efficiency (as a function of the system reduced gravitational potential, irradiation and planetary density) is given in the Appendix.

Notwithstanding the important limitation that our results only hold for atomic H+He atmospheres, they can be applied, for example, to perform systematic studies that investigate the role of photoevaporation-driven mass loss in shaping the observed distribution of planetary masses and radii. Additionally, they can be employed to compare the measured spectroscopic transit signatures (or the lack thereof) against realistic mass outflow rates; and to efficiently select prime targets for future transit spectroscopy campaigns. 


\begin{acknowledgements}
We thank the reviewer for their suggestions which have significantly improved
the content and clarity of the manuscript.
\end{acknowledgements}


\bibliographystyle{aa} 
\bibliography{bibliography.bib} 


\begin{appendix}
\section{Planet-dependent evaporation efficiency}

Based on the results presented in this work, we aim to provide an analytical approximation for the evaporation efficiency $\eta_{\rm eff}$ as a function of the relevant parameters at play. The ensuing best-fit efficiency is meant to replace the ``one-size-fits-all'' value that is routinely adopted within the framework of the energy-limited escape formalism, in the sense that it allows to employ the expression below regardless of whether the mass outflow rate is indeed energy-limited (according to the common definition):
  \begin{equation}
        \dot{M} = \eta_{\rm eff}\frac{3\flux}{4GK\rho_{\rm p}},
        \label{eq:EL_eta_eff}
    \end{equation}
where $F_{\rm XUV}$ is the XUV flux at the (average) orbital distance, $\rho_p$ is the mean planetary mass density and the factor $K$ accounts for the host star tidal forces\footnote{Following \cite{Erkaev2007}, the potential reduction factor $K$ can be expressed as $K=1-\frac{3}{2\xi}+\frac{1}{2\xi^3}$, where $\xi=R_{\rm RL}/R_p$ is the ratio between the Roche lobe radius and the planet radius.}.\\
For a direct application, the interested reader is directed to ATES online repository\footnote{\url{https://github.com/AndreaCaldiroli/ATES-Code}} where a dedicated script enables the straightforward estimate of the best-fitting $\eta_{\rm eff}$ (and ensuing $\dot{M}$) for a given planetary system. For completeness, we provide a quantitative description of the best-fitting analytical approximation in \S~\ref{sec:app1}. \\

\subsection{Analytical approximation}\label{sec:app1}
Noting that the energy-limited formalism implies a linear scaling of the mass loss rate with the irradiation to planetary density ratio, as well as the inverse of the $K$-factor, we aim to derive a best-fitting approximation as a function of those two variables. Concurrently, we wish to incorporate the dependence on planetary gravitational potential energy $\phi_p=GM_p/R_p$ highlighted by our analysis (\S~\ref{sec:phi}). 
Thus, for a given planetary system with known (i) planetary radius $R_p$ (defined as the optical transit radius); (ii) planetary mass, $M_p$; (iii) host stellar mass $M_\star$; (iv) stellar XUV flux $F_{\rm XUV}$ (see \S~\ref{sec:fluxappendix} below) measured at the (v) orbital distance $a$, we define the following quantities:
\begin{itemize}
\item
$\phired \equiv K\phi_p$ [erg g$^{-1}$] 
\item
$\phi_{\rm red,0}   = 10^{13.22}$ [{erg~g}$^{-1}$]
\item
$\tilde{F}_2 \equiv \frho$ in units of $10^2$ erg cm$^2$ s$^{-1}$ g$^{-1}$
\end{itemize}

The functional dependence of $\eta_{\rm eff}$ on $\phired$ and $\tilde{F}_2$ is chosen to reproduce the behavior of Figure \ref{fig:etaeff}, that is: at low-gravity ($\simlt$13.2, in log cgs units) $\eta_{\rm eff}$ is approximately constant ($\simeq \eta_0$) for a fixed value of $\tilde{F}_2$. At at high-gravity ($\simgt$13.2, in log cgs units), $\eta_{\rm eff}$ decreases sharply with increasing $\tilde{F}_2$, where the actual rate of decrease is in turn a function of $\tilde{F}_2$. 
The low- and high-gravity functional forms are joined through an appropriate weighting function, that is a sigmoid curve in the $\log\phired - \log\etaeff$ plane. 

The resulting functional shape is as follows: 
\begin{equation}
    \log \eta_{\rm eff} = A\, \phired^\alpha \sigma + \eta_0 (1-\sigma)
    \label{eq:eta_fit}
\end{equation}
with 
\begin{equation}
    \sigma\equiv  \left[1+\left(\frac{\phired}{\phi_{\rm red,0}}\right)^\beta\right]^{-1},\quad\beta<0,
    \label{eq:eta_sigmoid}
\end{equation}

where $A$, $\alpha$, $\eta_0$, and $\beta$ are expressed as power-law combinations of $\tilde{F}_2$ and $\log\tilde{F}_2$, as appropriate, through Equations \ref{eq:eta_coeffs1}-\ref{eq:eta_coeffs2}.
The numerical coefficients were evaluated through a least-squares fitting \textsc{Matlab} routine\footnote{\url{https://www.mathworks.com/products/curvefitting.html}} making use of 81 $\eta_{\rm eff}$ values resulting from the simulations discussed in \S~\ref{sec:phi} and \S~\ref{sec:slopes} of the Paper, yielding: 
\begin{align}
    \label{eq:eta_coeffs1}
    & A = 1.682\left(\log \tilde{F} _{2}\right)^{0.2802} - 5.488,\\[8pt]
    & \alpha = 0.02489\,\tilde{F} _2^{-0.0860} - 0.01007\,\tilde{F} _2^{-0.9543}, \\[8pt] 
    & \eta_0 = -0.03973\left(\log \tilde{F} _2\right)^{2.173} - 0.01359, \\[8pt] 
    & \beta  = - 0.01799\,\tilde{F}_2^{0.1723} - 3.3875\,\tilde{F} _2^{0.0140}. \label{eq:eta_coeffs2}
\end{align}

\begin{figure}
    \centering
    \includegraphics[width = \columnwidth]{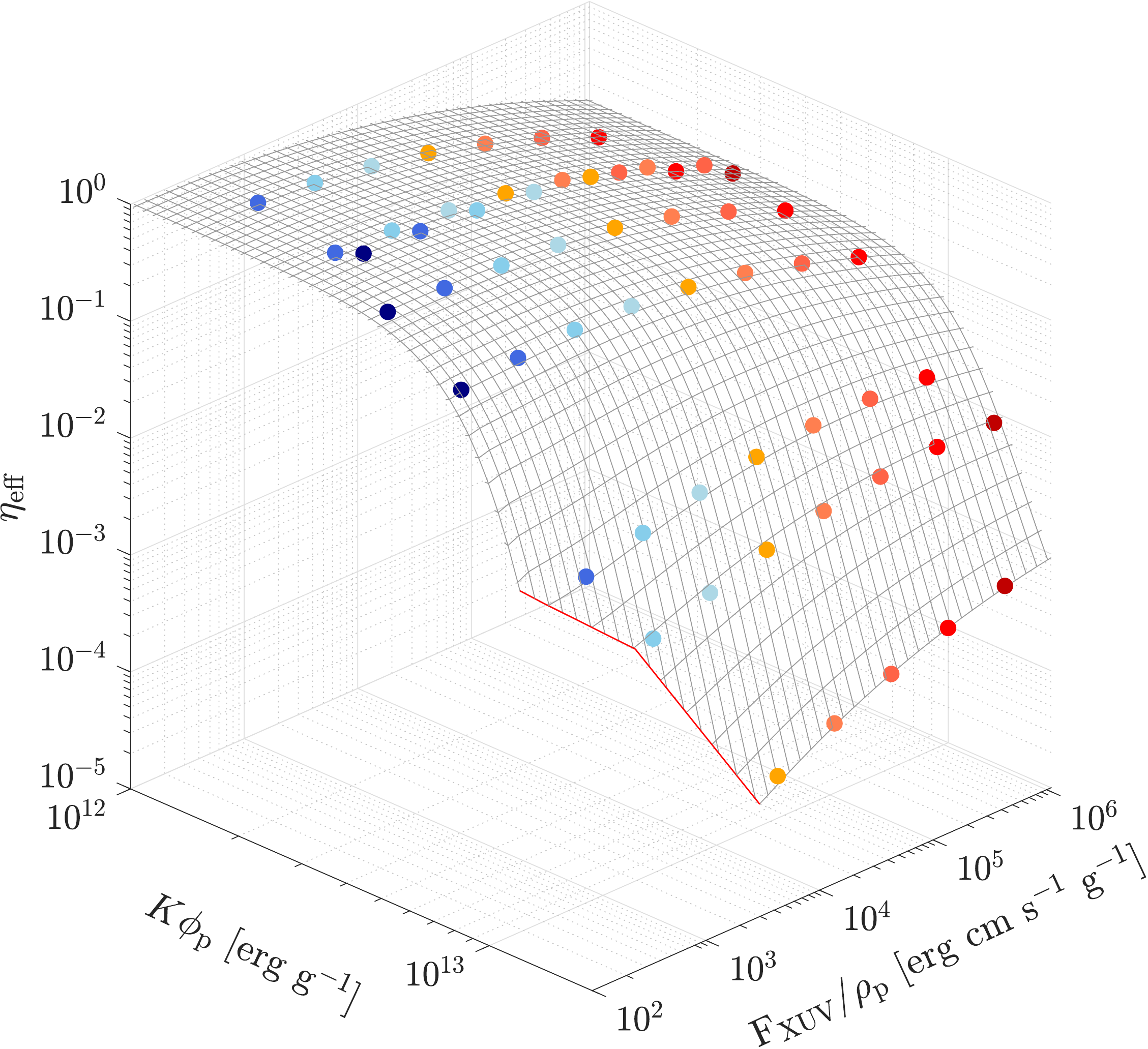}
    \caption{Analytic approximation of the function $\eta_{\rm eff}(\flux/\rho_p,\Kphi)$ as given by Equations~\ref{eq:eta_fit}-\ref{eq:eta_coeffs2}. The data points are color-coded to reflect the intensity of the XUV irradiation, according to the same color scheme as in Figure~\ref{fig:etaeff} where $F_{\rm XUV}$ varies between $10^2$ (deepest blue) and $10^6$ (deepest red) erg s$^{-1}$ cm$^{-2}$.  The thick red line represents the convergence limit of the ATES code described in \S 6.1 of \citet{ates1}.}
    \label{fig:fit1}
\end{figure}

\begin{figure}
    \centering
    \includegraphics[width = \columnwidth]{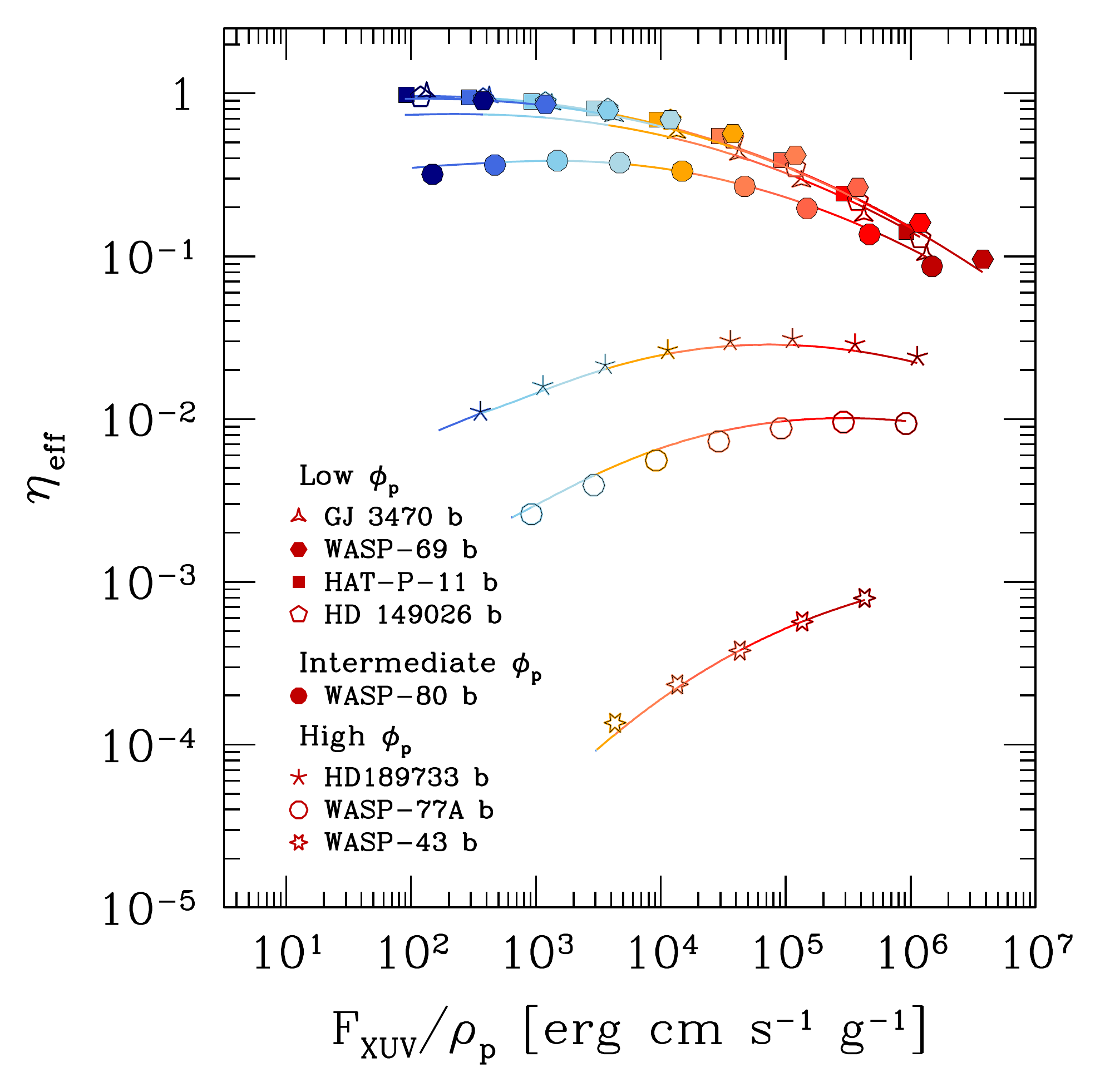}
    \caption{Projection of the analytical approximation illustrated in Figure \ref{fig:fit1} on the $\frho-\eta_{\rm eff}$ plane.}
    \label{fig:fit2}
\end{figure}

Figure~\ref{fig:fit1} illustrates the full functional dependence of the best-fitting $\eta_{\rm eff}$ on $F_{\rm XUV}/\rho_p$ and $K \phi$; its projection on the $\frho-\eta_{\rm eff}$ plane is shown in Figure~\ref{fig:fit2}.
The maximum discrepancy between the $\dot{M}$ values obtained by ATES (and discussed in \S~\ref{sec:phi} and \S~\ref{sec:slopes} of the paper) and the best-fit values obtained through Equation~\ref{eq:EL_eta_eff} is $\lesssim 0.16$ dex; the mean deviation is $\simeq 0.075$ dex. 
We remark that the above formulas are formally valid within the simulated range of parameters, that is for $10^2 \lesssim \frho \lesssim 10^6$ and $10^{12.17}\lesssim \Kphi \lesssim 10^{13.29}$ (in cgs units).


\subsection{$\flux$ estimate}\label{sec:fluxappendix}

The mass loss rates based upon which the above fit is derived are obtained by ATES under the assumption of a power-law-like photoionizing spectrum. Specifically, ATES assumes a piece-wise power law of the form $F \propto E^{-1}$, both in the EUV ($[13.6-124]$ eV) and in the X-ray band ($[0.124-12.4]$ keV). In each band, the spectrum is normalized to the (user-provided) input EUV and X-ray luminosity values. 
As discussed in Section 6 of \cite{ates1}, this choice is motivated by the fact that the steady-state mass loss rates mainly depend on the total number of photoionizing photons, rather than the chosen stellar spectral shape. 

\end{appendix}

\end{document}